\documentclass[12pt]{iopart}
\usepackage{graphicx}
\usepackage{url}
\usepackage{cite}
\usepackage{palatino}

\newlength{\shorter}
\begin{document} % here for iopart

\date{26 September 2016}

%%%%%%%%%%%%%%%%%%%%%%%%%%%%%%%%%%%%%%%%%%Title%%%%%%%%%%%%%%%%%%%%%%%%%%%%%%%%%
\title[Nano-scale thermal transfer]{Nano-scale thermal transfer\\
 -- an invitation
to fluctuation electrodynamics}
\author{Carsten Henkel$^*$%
\footnote[1]{henkel@uni-potsdam.de}%
}
\address{$^*$\ Institute of Physics and Astronomy, University of Potsdam,\\
Karl-Liebknecht-Str. 24/25, 14476 Potsdam, Germany}

%%%%%%%%%%%%%%%%%%%%%%Abstract%%%%%%%%%%%%%%%%%%%%%%%
\begin{abstract}
An electromagnetic theory of thermal radiation is outlined,
based on the fluctuation electrodynamics of Rytov and
co-workers. We discuss the basic concepts, the status of
different approximations. The physical content is illustrated
with a few examples on near-field heat transfer.
\end{abstract}
%%%%%%%%%%%%%%%%%%%%%%%%%%%%%%%%%%%%%%%%%%%%%%%%%%%%%


\submitto{Zeitschrift f\"ur Naturforschung A, special issue ``Heat Transfer and Heat Conduction on the Nanoscale'' edited by S.-A. Biehs,
P. Ben-Abdallah, and A. Kittel}
%%%%%%%%%%%%%%%%%%%%%%%%%%%%%%%%%%%%%%%%%%%%%%%%%%%%%%%%%%%%%%%%

\section*{Introduction}

\subsection*{Motivation}

The papers in this special issue illustrate that for the
management of heat and energy transfer in nano-technology,
we are awaiting fundamental challenges. It may suffice to mention
Moore's law according to which on a scale of 
five to ten years, the microscopic structure of semiconductor
junctions will become relevant. We will have to cope with
statistical and thermal fluctuations 
that are significant compared to nominally specified 
(mean) values. At the same time, it is likely that the design
of novel materials with taylored properties in photon and
phonon transport opens up new avenues like `thermal computing'
and raises the efficiency of thermoelectric and thermo-photovoltaic 
devices. 

In this paper, we intend to provide a gentle introduction into
a statistical description of thermal radiation and radiative
heat transfer 
that may guide the engineer from the micro-scale 
down to the nano-scale. 
The key tool is already more than 60 years old, it has been
dubbed fluctuation electrodynamics and was developed in the
1950s by physicists in the former Soviet Union: 
Sergei M. Rytov~\cite{Rytov3}, Vladimir L. Ginzburg~\cite{Ginsburg53}, 
I. E. Dzyaloshinskii, E. M. Lifshitz, 
and L. P. Pitaevskii~\cite{Lifshitz61,Landau9},
just to mention a few. A selection of more recent reviews are
Refs.\cite{BenAbdallah15, Song15b, Liu15b, Jones13, Dorofeyev11b, Narayanaswamy09, Greffet07b, Henkel04c, Henry96}. The main idea may be explained by analogy to
Brownian motion: instead of characterizing the motion of 
a particle by its thermal energy, one introduces microscopic
trajectories that are perturbed by randomly fluctuating forces.
The forces arise from the interaction to a thermal environment, 
for example between a liquid and a colloidal particle immersed in it. Since the seminal work of Perrin, Langevin and 
Einstein~\cite{Einstein05b}, it has become clear that an
ensemble of such fluctuating trajectories can
recover (or `unravel') thermal equilibrium statistics 
and may even describe systems
driven out of equilibrium, for example by temperature gradients
or external forces. 

The radiation field coupled to matter at finite temperature
may be described in a similar way. Fluctuation electrodynamics
thus takes serious Planck's idea that thermal equilibrium for
blackbody radiation arises from the exchange of energy with
the `cavity walls'. The latter realize a macroscopic material 
system whose `internal temperature' imposes the steady state properties of the radiation field. The field is indeed unable 
to thermalize by itself (excluding extreme situations like 
the relativistic electron-positron plasma during the first minutes
of the Universe). 
The equations of fluctuation electrodynamics are involving 
in an essential way
the response of the material: dispersion and absorption. It is
well known since the 19th century that this can be modelled 
using material parameters
like permittivity, permeability, and conductivity. The main
insight of the current trend towards the nano-scale is that
this `macroscopic description' can be extended in a quite
natural way to scales smaller than the typical wavelengths
of thermal radiation (around $1\,\mu{\rm m}$, say). There is
a relatively large window of scales (the realm of 
nano-technology) where an atomistic description of matter
is not yet needed -- it may be dubbed the `mesoscopic range'.
Let us compute as a rough estimate the ratio between the
radiation and matter degrees of freedom. Fixing a cube 
of size $L^3$ of a medium with refractive index $n$, say,
the number of photonic modes in a frequency band is
\begin{equation}
\rho_{\rm rad} \, {\rm d}\omega 
%L^3 d^3k = 1/(2pi)^3
%4pi solid angle
%2 polarization
%k = omega/(c n)
\sim \frac{ \omega^2 L^3 }{ (n c)^3 }
\, {\rm d}\omega 
\,,
\label{eq:photon-mode-density}
\end{equation}
while the microscopic degrees of freedom are given by the 
atomic number density $\varrho / m$
\begin{equation}
\rho_{\rm mat} \, {\rm d}\omega 
\sim \frac{ \varrho L^3 }{ m \omega_D } \, {\rm d}\omega 
\,.
\label{eq:matter-mode-density}
\end{equation}
We have assumed for simplicity that the phonon modes
are distributed evenly up to the Debye frequency $\omega_D$.
Using estimates for a cubic crystal, the photonic mode density
is much less by a small factor $\sim (v/c) (a/\lambda)^2 \ll 1$
where $v$ is the speed 
of sound, $a$ the unit cell size, and $\lambda$ the 
photonic wavelength. This illustrates that matter can indeed
play the role of a macroscopic reservoir 
for the radiation field. 

For the purposes of this review, we focus on the simplest
macroscopic electrodynamics where the material response is
assumed to happen locally. Ohm's law, for example, is written
${\bf j} = \sigma {\bf E}$ where all quantities are evaluated
at the same $({\bf r}, \omega)$. In metals, this approximation 
is expected to work on spatial scales longer 
than the mean free path of charge carriers
and the Fermi wavelength. To the same level of approximation,
interfaces between materials are assumed to be `sharp'. The
permittivity $\varepsilon( {\bf r}, \omega )$, for example, 
varies like a step function across the boundary. This is 
accompanied by
suitable boundary conditions for the macroscopic fields.
Descriptions that go beyond this picture would use spatial
dispersion (non-local response) and genuine response functions
for the interface region like a surface current or 
a surface polarization~\cite{GarciaMolinerBook, BedeauxVlieger, Liebsch93}. 
A discussion of how to formulate
fluctuation electrodynamics in nonlocal media can be found
in Refs.\cite{AsgerMortensen14,Buhmann12b,Pitaevskii09,Barash75}.

The last approximation that is commonly applied is the linearity
of the medium response. For material parameters like
$\varepsilon$, linear response is written into its definition.
The physics beyond this regime is very rich and contains effects
like rectification of thermal noise, generation of higher 
harmonics, temperature-dependent material parameters, hysteresis, 
and so on. Some of these aspects are already studied in
nano-scale thermal devices, in the proposals for thermal
diodes or rectifiers, for example. Others are being explored
theoretically~\cite{Scheel06b, Soo16}.
From a mathematical viewpoint,
the linear approximation provides a simple link between the
probability distributions of the Langevin forces (matter-related
sources of the field) and of the field, yielding eventually
to Gaussian statistics which is completely characterized
by mean values and the correlation spectra. 

%continuum / step-like description of matter,
%length scales (`local' and `beyond'),
%linearity (gaussian approximation)

%system (= field + part of matter) + bath (damping of matter
%polarization), compare mode densities (photon vs phonon).

A number of recent experiments on nano-scale heat transfer
can be found in this Special Issue, one of the oldest we are
aware of is from the Dransfeld group~\cite{Xu94}. We shall comment on
a few recent references in this paper. As a side remark, we mention
techniques where thermal dynamics becomes accessible on
short time scales, comparable to the equilibration time between
between electron and phonons~\cite{Shayduk11}. 
They may provide a complementary approach 
to a better understanding of heat transport, where the
transient response of matter is in the focus.
 
\begin{figure}[htb]
\centerline{%
\includegraphics*[width=120mm]{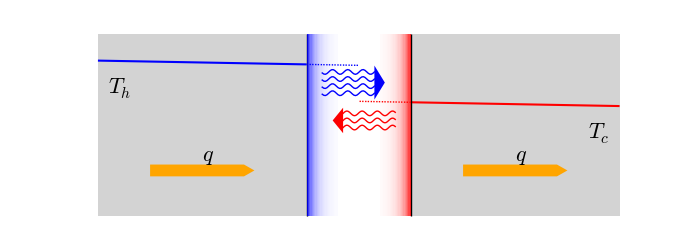}}
\caption[]{Typical setting of a heat transfer experiment
on the nano-scale. The blue and red lines give the temperature
profile in the two bodies (left: hot, right: cold). The heat
current $q$ is carried by conduction in the body and by
the difference of thermal emission in the gap. The blue and
red shadings illustrate the evanescent waves in the near-field
of the surfaces; they provide
additional `tunnelling channels' for heat transfer. This can
lead to an amplification by a few orders of magnitude above
the Stefan--Boltzmann law.
}
\label{fig:sketch-ngap}
\end{figure}

The sketch in Fig.\ref{fig:sketch-ngap} below is meant to
fix some ideas: two bodies are described as extended objects
with sharp boundaries. Inside, a local temperature can be
defined, although its variation may be weak despite a
sizeable heat current. The bodies are separated by a vacuum 
gap across which photons are `propagating' and `tunnelling' 
to transport energy. Direct contact which would lead to phononic 
and
electronic transport, can be excluded by working at distances
above a few \AA{}ngstrom.

\section*{Rytov formulation of fluctuation electrodynamics}

\subsection*{Maxwell-Langevin equations}

Within the approximations outlined in the Introduction,
the radiation field in a mesoscopic medium is described
by the following set of macroscopic Maxwell equations
\begin{eqnarray}
\nabla \cdot \varepsilon {\bf E} = \rho
\,,\qquad
&&
\nabla \times {\bf H} - {\rm i}\omega \varepsilon {\bf E} = {\bf j}
\nonumber
\\
\nabla \cdot \mu {\bf H} = 0
\,,\qquad
&&
\nabla \times {\bf E} - {\rm i}\omega \mu {\bf H} = {\bf 0}
\label{eq:meso-Maxwell}
\end{eqnarray}
Here, $\varepsilon = \varepsilon( {\bf r}, \omega )$ 
and $\mu$ describe the electric and magnetic response of the 
medium.
The charge and current densities can be represented
by (`external') polarization and magnetization fields
\begin{equation}
\rho = - \nabla \cdot {\bf P}
\,,
\qquad
{\bf j} = - {\rm i}\omega {\bf P} + \nabla \times {\bf M}
\label{eq:rho-and-j}
\end{equation}
In the case of non-magnetic and local conductors, 
the key response function
is the conductivity $\sigma$, and Eqs.(\ref{eq:meso-Maxwell})
apply with $\varepsilon = \varepsilon_0 - {\rm i} \sigma/\omega$,
$\mu = \mu_0$, and ${\bf M} = {\bf 0}$. In some applications
to thermal radiation in metals, a `background' dielectric response 
(due to the ionic cores)
is taken into account, replacing $\varepsilon_0$ by $\varepsilon_b$
inside the conductor.

The boundary conditions for the fields at a smooth interface
follow from the Maxwell equations~(\ref{eq:meso-Maxwell}) 
when understood in the sense
of distribution theory: a jump in the normal component of the
displacement field $\varepsilon {\bf E}$, for example, is given
by the surface charge density, i.e., 
a $\delta$-sheet of $\rho( {\bf r} )$,
localized at the interface. Similar techniques may be applied
when the response functions vary rapidly near the interface on
the spatial scales relevant for the fields. One may introduce
a `macroscopic' model for the interface where the material equations
are extrapolated from the two bulk media to the surface. The
deviations from this extrapolation are called `excess charges'
or currents and are represented by $\delta$-distributions localized
on the (macroscopic, idealized) interface. For details, see the
books~\cite{GarciaMolinerBook,BedeauxVlieger}.

The key idea of fluctuation electrodynamics, developed by
Rytov and co-workers~\cite{Rytov3}, is that the sources
$\rho$, ${\bf j}$ are random or fluctuating fields. This is
why we shall call Eqs.(\ref{eq:meso-Maxwell}) the Maxwell-Langevin
equations. If the
sources are nonzero on average, their values would be interpreted 
as `external'
charges as in the conventional macroscopic electrodynamics. The
fluctuations around the average arise from the thermal motion
of carriers in the medium and are therefore determined by its
temperature and its `oscillator strength' (or density of states).
Within the macroscopic scheme, as explained above, it is
natural to apply local thermodynamic equilibrium statistics
to describe the charge and current fluctuations.

\subsection*{Source (current) fluctuations}

To illustrate these concepts, let us take as
an example the current fluctuations in a conductor and assume
for simplicity that the mean value 
$\langle {\bf j}( {\bf r}, \omega ) \rangle = {\bf 0}$.
The fluctuations of the current
define its noise spectrum as follows
\begin{equation}
\langle j_m^*( {\bf r}', \omega' )
j_n( {\bf r}, \omega ) \rangle = 
2\pi \delta( \omega' - \omega )
\langle j_m( {\bf r}' )
j_n( {\bf r} ) \rangle_{\omega}
\label{eq:def-jj-spectrum}
\end{equation}
The writing $\langle \ldots \rangle_{\omega}$
on the rhs is adopted to avoid additional notation, although it
is, strictly speaking, slightly abusive. This quantity
provides the (power) spectral density,
i.e. the Fourier transform of the autocorrelation function
(Wiener-Khintchin formula)
\begin{equation}
\langle j_m( {\bf r}', t' ) j_n( {\bf r}, t ) \rangle
= \int\limits_{-\infty}^{\infty}\!\frac{ {\rm d}\omega }{ 2\pi }
{\rm e}^{ {\rm i} \omega( t' - t ) } 
\langle j_m( {\bf r}' )
j_n( {\bf r} ) \rangle_{\omega}
\label{eq:Wiener-Khintchine}
\end{equation}
Note that the factor $2\pi$ in Eq.(\ref{eq:def-jj-spectrum}) 
is tied to this convention for the Fourier transform and that
such a definition makes sense in a stationary situation only 
(correlations only depend on the time difference). The complex
Fourier expansion used here is appropriate for `quantum
detectors' like a photomultiplier. Glauber's theory of these
detectors works with the hermitean field operator 
${\bf E}( {\bf r}, t )$ and shows that the signal involves 
the convolution of ${\bf E}$ with a complex exponential
${\rm e}^{ - {\rm i} \omega t }$ where $\omega > 0$ is the
threshold frequency (work function). In classical electronics,
the signal is convolved with a real-valued reference or circuit
response. In that
case, an autocorrelation function based on a symmetrized operator 
product is relevant. Its spectrum is symmetric and is equal
to the average of its
`quantum cousin'~(\ref{eq:Wiener-Khintchine}) over positive and negative frequencies. 

Thermal equilibrium and local macroscopic electrodynamics
yield the following relation between spectrum and conductivity
\begin{equation}
\langle j_m( {\bf r}' )
j_n( {\bf r} ) \rangle_{\omega}
= 2\hbar 
\frac{ \omega \mathop{\rm Re}\sigma( \omega ) }{ {\rm e}^{ \hbar \omega / k_B T } - 1 }
\,\delta_{mn} \delta( {\bf r}' - {\bf r} )
\equiv
S_j( {\bf r}, \omega ) 
\,\delta_{mn} \delta( {\bf r}' - {\bf r} )
\label{eq:FDr-current}
\end{equation}
where both $T = T( {\bf r} )$ and $\sigma$ 
may vary spatially (local temperature).
\begin{figure}[htb]
\centerline{%
\includegraphics*[width=120mm]{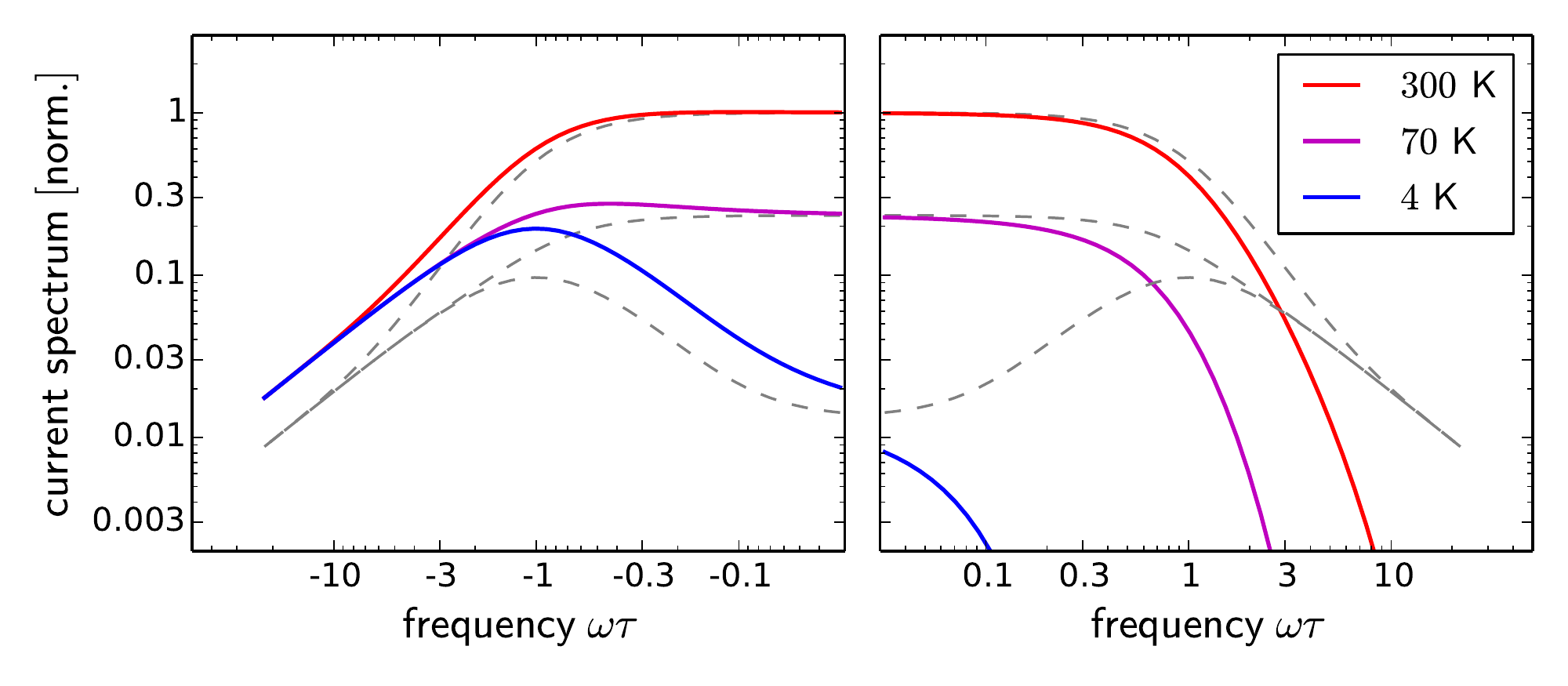}}
\caption[]{Spectral density $S_j( \omega )$
of current fluctuations inside
a Drude conductor, Eq.(\ref{eq:FDr-current}), for positive
and negative frequencies. 
The spectrum
is normalized to its low-frequency limit at $T = 300\,{\rm K}$
[see Eq.(\ref{eq:Nyquist})],
the temperature dependence of $\sigma_{\rm DC}$ is neglected
for simplicity.
The frequency is normalized to the Drude relaxation rate 
$1/\tau$ (we took $\hbar/\tau = 10\,{\rm meV}$).
The dashed lines give the spectrum for the 
symmetrized autocorrelation of the current. 
}
\label{fig:current-spectrum}
\end{figure}
The real part (the conductance) of the complex admittance
$\sigma$ is related to Ohmic dissipation, hence the name 
fluctuation--dissipation (FD) relation for 
Eq.(\ref{eq:FDr-current}). 
This formula is the quantum extension of Johnson--Nyquist
noise, see the seminal paper by Callen and Welton on the
FD theorem~\cite{Callen51} and the review by Ginzburg~%
\cite{Ginsburg53}. 
The current spectrum $S_j( {\bf r}, \omega )$
is plotted in Fig.\ref{fig:current-spectrum} inside a 
Drude conductor, $\sigma( \omega ) = \sigma_{\rm DC}
/(1 + {\rm i} \omega \tau)$. The plot covers both
positive and negative frequencies: note the asymmetry of
the spectrum. This is typical for operator products that are
not in symmetric order and hence a typical quantum effect.
Indeed, with our convention for the spectral density, quantum
(or vacuum) fluctuations become visible at negative
frequencies. The scheme required to detection these 
would be based not
on absorption, but on emission (spontaneous and stimulated),
see for example Ref.\cite{Usami04a}.
The quantum current fluctuations in a Drude conductor
show a broad Debye-like peak at $\omega \sim - 1/\tau$ where
$\tau$ is the electronic relaxation time. (The mean free path
is $v_F \tau$ with $v_F$ the Fermi velocity.) Another convention
is to take a symmetrized correlation function so that the
spectrum becomes symmetric (gray dashed lines).

At lower frequencies,
the current noise is white and given by the Nyquist formula,
\begin{equation}
\langle j_m( {\bf r}' )
j_n( {\bf r} ) \rangle_{\omega}
\approx 
2 k_B T \sigma_{\rm DC}
\,\delta_{mn} \delta( {\bf r}' - {\bf r} )
\label{eq:Nyquist}
\end{equation}
It is interesting to note that for typical
Drude parameters and room temperature, the `thermal energy-time 
uncertainty product', $k_B T \tau / \hbar$, is not far from unity.
In other words, good conductors show strongly damped
electrical currents at room temperature.

\subsection*{Field correlations}

The thermal radiation field generated by the Rytov-Langevin
Eqs.(\ref{eq:meso-Maxwell}) depends linearly on its sources.
Solutions to the homogeneous equations can be used to describe
radiation incident from infinity~\cite{Savasta00}. 
Some authors prefer to
cancel these solutions by allowing for a nonzero imaginary
part in $\varepsilon( {\bf r}, \omega )$ in all 
space (see, for example, Ref.\cite{Scheel00b}). We follow
this route and get for bodies in local equilibrium the
following expression for the electric field correlation
function (defined as in Eq.(\ref{eq:def-jj-spectrum}),
summation over double indices):
\begin{equation}
\langle E_k( {\bf r} ) E_l( {\bf r}' ) \rangle_\omega
=
\int\!{\rm d}^3s \,
G_{km}^*( {\bf r}, {\bf s}, \omega )
G_{lm}( {\bf r}', {\bf s}, \omega )
S_j( {\bf s}, \omega )
\label{eq:fields-from-sources}
\end{equation}
Here $G_{kl}( {\bf r}, {\bf s}, \omega )$ is the
electromagnetic Green tensor which gives the electric and
magnetic fields
radiated by a point current source located at ${\bf s}$:
\begin{eqnarray}
E_k( {\bf r}, \omega ) &=& 
\int\!{\rm d}^3s\, 
G_{kl}( {\bf r}, {\bf s}, \omega ) j_l( {\bf s}, \omega )
\label{eq:def-Green-tensor}
\\
H_k( {\bf r}, \omega ) &=& 
\frac{ - {\rm i} }{ \omega \mu( {\bf r}, \omega ) }
\epsilon_{klm}
\frac{ \partial }{ \partial r_l }
\int\!{\rm d}^3s\, 
G_{mn}( {\bf r}, {\bf s}, \omega ) j_n( {\bf s}, \omega )
\label{eq:def-H-Green-tensor}
\end{eqnarray}
As a function of ${\bf r}$ and the index $k$, $G_{kl}$ 
satisfies the boundary conditions at macroscopic interfaces.
Many different conventions for prefactors in 
Eqs.(\ref{eq:def-Green-tensor}, \ref{eq:def-H-Green-tensor}) 
appear in the literature.
Explicit expressions are available in free space~\cite{Jackson}, 
at a planar interface~\cite{Wylie84,Sipe87} and in 
multi-layer structures~\cite{Tomas95}, in spherical~\cite{Chew87} 
and cylindrical geometries~\cite{ParsegianBook}. The book
by Chew~\cite{ChewBook} provides an exhaustive discussion,
including numerical techniques from computational electrodynamics.
Let us simply note that in practice, it is
possible to avoid the volume integration in 
Eq.(\ref{eq:fields-from-sources}) and get instead an integral
over the surfaces of the bodies~\cite{Levin74,Dorofeyev11b,Reid13}.
(The price to pay is the
restriction to spatially constant temperatures inside the 
bodies.) For objects with highly symmetric shape, an expansion
into orthogonal vector field modes is possible (`principle
of sufficient symmetry'), reducing Eq.(\ref{eq:fields-from-sources})
to a summation over products of mode functions.

From a correlation function like 
$\langle E_k( {\bf r} ) E_l( {\bf r}' ) \rangle_\omega$, it is
easy to compute the electromagnetic energy density (take $k=l$
and ${\bf r} = {\bf r}'$), correlations for the magnetic field
(take the curl with respect to both ${\bf r}$ and ${\bf r}'$),
and the Poyting vector, for example. All quantities naturally
appear in the spectral domain, and can be split into contributions
originating from the different objects. In this way, heat currents 
from body $A$ to $B$ and back can be determined by computing the 
flux of the Poynting vector across a surface that separates $A$ and 
$B$ [see Eq.(\ref{eq:Polder71-flux}) below]. 
The elements of the electromagnetic stress 
tensor~\cite{Jackson}, plotted
locally, permit to visualize forces (momentum transfer) between
objects, providing physical insight into dispersion and Casimir
forces; see Ref.~\cite{Rodriguez07} for examples. A seminal
example of this problem is Lifshitz' derivation of the 
van der Waals force between macroscopic bodies~\cite{Lifshitz56, Lifshitz61}.

\subsection*{Quantum field theory}

The Maxwell-Langevin approach sketched so far provides the
equations of motion of the field (operators), and relevant 
averages (correlation functions) are obtained from those of
the source currents. The question has been raised what would
be the Hamiltonian of this theory. The quantization scheme
for the macroscopic Maxwell equations developed by the Kn\"oll
and Welsch group in Jena~\cite{Scheel00b,BuhmannBookI} 
gives an answer in line
with the ideas of Rytov and co-workers~\cite{Rytov3}: the
information contained in the macroscopic medium response
functions (dielectric function $\varepsilon$, \ldots) is sufficient.
We adopt 
here a slight re-writing of the Jena equations
and keep working with the current operators
${\bf j}( {\bf r}, \omega )$. The basic idea is to upgrade
each Fourier component to a dynamic variable. The medium+field
Hamiltonian is then given by the sources alone
\begin{equation}
H_{\rm MF} = \int\limits_{0}^{\infty}\!\frac{ {\rm d}\omega }{ 2\pi }
\int\!{\rm d}^3r\,
\frac{ {\bf j}^*( {\bf r}, \omega ) \cdot
{\bf j}( {\bf r}, \omega ) }{ 2 \mathop{\rm Re}
\sigma( \omega ) }
\label{eq:Hamiltonian-current}
\end{equation}
It is interesting
that the power density of Joule absorption provides the
energy~(\ref{eq:Hamiltonian-current}) of medium and field.
The Heisenberg equations of motion yield the seemingly
trivial time evolution $\sim {\rm e}^{ - {\rm i} \omega t }$
provided the commutator of the currents is chosen as
\begin{equation}
\left[ j_m( {\bf r}', \omega' )
\,,
j_n^*( {\bf r}, \omega ) \right]
= 4\pi 
\hbar \omega \mathop{\rm Re}\sigma( \omega )
\,\delta_{mn} \delta( {\bf r}' - {\bf r} )
\delta( \omega' - \omega )
\label{eq:commutator-current}
\end{equation}
The relation ${\bf j}( {\bf r}, \omega ) 
= {\bf j}^*( {\bf r}, -\omega )$ is imposed as initial 
condition and is preserved during the evolution. 
The local thermal equilibrium ensemble
is generated by a density operator proportional to the exponential
of $H_{\rm MF}$, taking the local temperature field $T({\bf r})$
under the spatial integral in Eq.(\ref{eq:Hamiltonian-current}).
The FD relation with its correlation function~(\ref{eq:FDr-current})
then follows.

The
observable electric field is given by the source current
convolved with the Green function~[Eq.(\ref{eq:def-Green-tensor})],
and this relation can be understood as a generalized mode expansion.
In this way, the Maxwell-Langevin equations are satisfied.
The coupling to external sources (atoms or molecules) 
proceeds in the usual way by adding interaction terms to
$H_{\rm MF}$. A multipolar coupling scheme
is quite natural, for details see the book by Buhmann~%
\cite{BuhmannBookI}.
An important consistency check is to recover the Pauli-Jordan
commutator between the electric and magnetic fields:
\begin{equation}
\left[
E_i( {\bf r}, t )
, \,
B_j( {\bf r}', t )
\right] = 
- \frac{ {\rm i} \hbar }{ \varepsilon_0 }
\epsilon_{ijk} \frac{ \partial }{ \partial x_k }
\delta( {\bf r} - {\bf r}' )
\label{eq:}
\end{equation}
This is achieved by using Eq.(\ref{eq:commutator-current}) 
for the source currents, the Kramers-Kronig relations for
the Green function, and the following
identity~\cite{Eckhardt84,Scheel98a}
\begin{equation}
\int\!{\rm d}^3s
\mathop{\rm Re}(\sigma)\,
G^*_{ik}( {\bf r}, {\bf s} )
G_{jk}( {\bf r}', {\bf s} )
=
- \mathop{\rm Re}G_{ij}( {\bf r}, {\bf r}' )
\label{eq:}
\end{equation}
that follows from the classical macroscopic Maxwell equations.
The frequency arguments were suppressed for simplicity,
and the negative real part appearing on the rhs is due to the
convention adopted here for the Green tensor [compare to the
energy conservation law~(\ref{eq:power-conservation}) below].

\subsection*{Discussion}

Let us finish this basic introduction with a remark on
physical interpretation.
A cursory scan through the literature yields different answers
as to the actual status of the current fluctuations appearing
in Eqs.(\ref{eq:meso-Maxwell}, \ref{eq:FDr-current}). One may
consider them as \emph{mathematical artefacts} whose only merit is to
reproduce, in thermal equilibrium, the spectrum of the
radiation field~\cite{Agarwal75a}, similar to thermostatting
Langevin forces in molecular dynamics.
It has been checked that equilibrium is indeed recovered 
if all bodies are at the same temperature: the field then 
follows Planck's blackbody spectrum, modified by the bodies' 
emissivity and enhanced significantly when near-field components 
become detectable at sub-wavelength distances~\cite{Agarwal75a}. 
Right from its inception, however, 
fluctuation electrodynamics has been applied
to non-equilibrium settings, too~\cite{Landau9}, where this
reference situation is no longer available. One is then
inclined to adopt the viewpoint that Rytov's fluctuating currents
are a model for \emph{actual thermal fluctuations} of matter
observables that are relevant as electromagnetic sources---%
an approximate model, of course, since in the formulation
presented above, the local macroscopic theory has been applied.
But the model has some reasonable footing given the large
number of microscopic constituents that make up a condensed-matter
system on a mesoscopic length scale of, say, a few nanometers
[see Eq.(\ref{eq:matter-mode-density})].
The key assumption is, of course, that on this scale, the concept
of a local temperature $T( {\bf r} )$ makes sense, and that the
electromagnetically relevant quantities have fluctuations that
are equilibrated at this temperature. Such a model is eventually
required in any description, as described insightfully by
Barton~\cite{Barton15}:
\begin{quote}\small
One could of course try to pursue the further question how the Langevin forces themselves might be kept functioning as envisaged; to explore this would then entail hypothesising some thermostats controlling the forces, and calculating the forces instead of making assumptions about them a priori; and so on, potentially ad infinitum. In the end, one must necessarily settle for control at some level through thermostats that are external to the system, in the sense that they impose
temperatures by fiat, through dynamics that do not enter the calculation and are not spelled out.
\end{quote}
The following formula may illustrate the problem of local 
equilibrium
\begin{equation}
{\bf j}_{\rm tot}( {\bf r}, \omega) = 
\sigma {\bf E}( {\bf r}, \omega) +
{\bf j}( {\bf r}, \omega)
\label{eq:Rytov-split}
\end{equation}
This sum (sometimes called the `Rytov split') gives the current 
as the sum of Ohm's law (`induced current'), 
while the second term gives the fluctuations around this
value (`fluctuating current', sometimes marked with subscript
${\rm fl}$). In a non-equilibrium setting like in heat transfer,
the field ${\bf E}( {\bf r}, \omega)$ is in general not
in a state of local equilibrium because it contains radiation 
from distant sources at different temperatures. The 
current is then not in local equilibrium neither, because its
induced part is
simply the linear (and local) response of the medium to this 
non-equilibrium field. But the fluctuations around it
(fluctuating current) are characterized by the local temperature.
This has been recognized as the key assumption of fluctuation
electrodynamics, 
for example by Barton~\cite{Barton15,Barton16} who comments
in Ref.\cite{Barton15}
on Loomis and Maris~\cite{Loomis94}:
\begin{quote}\small
[\ldots] who do not assume local thermal equilibrium, but only that in each half-space the noise is appropriate to the temperature of its thermostat.
\end{quote}
It is, of course, another problem how local thermal equilibrium
is established in the material, and whether this happens fast
enough to talk of a single temperature for all frequencies
of the local current noise. A practically-minded engineer might
want to use an equation of motion for the temperature distribution
$T( {\bf r} )$ using a combination of heat conduction and
heating/cooling by absorption/emission of thermal radiation.
It seems quite obvious that such a `self-consistent' model requires
a separation of time scales between the `fast' electromagnetic
fluctuations and the `slow' evolution of the temperature field.
Needless to say that for many experiments on near-field radiative
transfer, it is sufficient to
assume a stationary temperature profile, which may even be spatially
flat (across each body) if the thermal conductivity is large enough.
For a counter-example on short time scales, see the 
experiment of the Bargheer group quoted above~\cite{Shayduk11}.

As a quick estimate, let us consider a typical nano-scale heat 
flux that aims at
competing with the solar constant, $q = 1360\,{\rm W/m}^2$.
Typical thermal conductivities of condensed matter are in the range
$\kappa = 1\ldots 10^2\,{\rm W/K\,m}$ at room temperature.
On the spatial scale $L$ of a microdevice, one thus estimates
a temperature difference
\begin{equation}
\Delta T \sim L \frac{ \Delta T }{ L }
\sim L \frac{ q }{ \kappa } \sim 10^{-5} \ldots 10^{-3}\,{\rm K} \times ( L / 1\,\mu{\rm m} )
\label{eq:}
\end{equation}
% $L = 1\,\mu{\rm m}$,
which is probably below the precision of typical thermometry. 
In a nano-junction,
the temperature profile is thus `step-like' 
(see Fig.\ref{fig:sketch-ngap}). The radiation field in the gap
between the bodies has no well-defined temperature, of course. 
This is similar to radiation in the upper atmosphere which is 
a superposition of solar and terrestrial sources.

\section*{Examples}

We now present a few examples of results from fluctuation 
electrodynamics. Two viewpoints will be stressed:
the differences to radiative transfer which describes thermal
radiation on larger scales~\cite{Chandrasekhar},
and challenges beyond the mesoscopic theory presented so far.

\subsection*{Typical features on the nano-scale}

Thermal radiation is a classical topic in astrophysics
where it determines the luminosity and the inner structure
of stars. The cosmic
microwave background also provides an example where the
blackbody spectrum is a very accurate description.

\begin{figure}[htb]
\centerline{%
\includegraphics*[width=90mm]{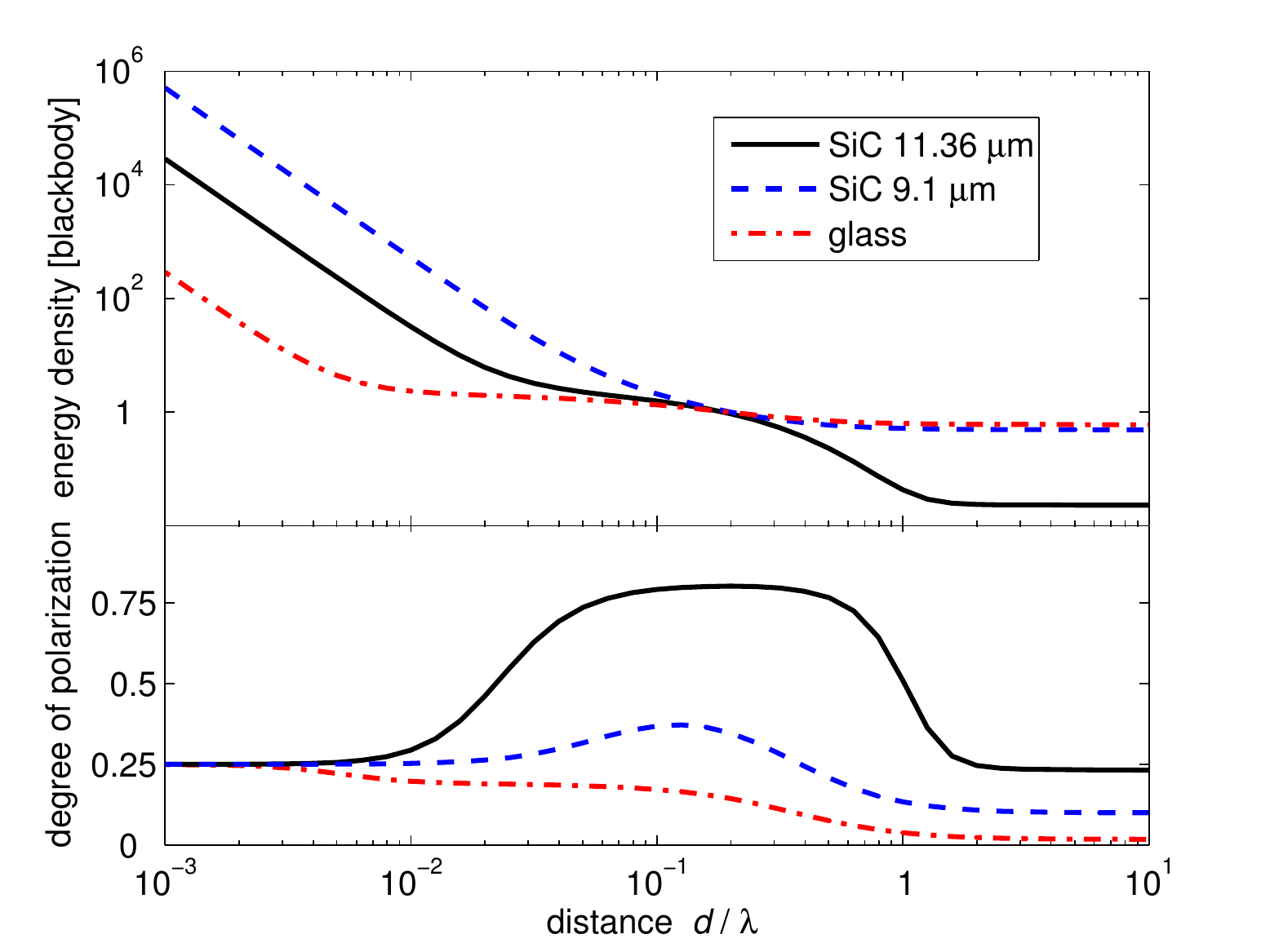}}
\caption[]{Electric energy density spectrum as a function of 
distance (top panel) and degree of polarization (bottom panel).
Parameters for SiC: 
$\varepsilon( 11.36\,\mu{\rm m} ) = -7.6 + 0.4\,{\rm i}$ 
and
$\varepsilon( 9.1\,\mu{\rm m} ) = 1.8 + 4.0\,{\rm i}$;
for glass
$\varepsilon( 500\,{\rm nm} ) = 2.25 + 10^{-3}\,{\rm i}$.
The energy densities follow a power law $1/d^2$ at short 
distances. They have been normalized to the Planck spectrum.
Figure adapted from Ref.\cite{Greffet07b}, Fig.~8.}
\label{fig:SiC-glass}
\end{figure}

As one works with condensed matter sources, 
deviations from the Planck spectrum become apparent. Some of
these are well-known and are described by the classical
Kirchhoff concepts of emissivity and absorption 
(`grey bodies')~\cite{Chandrasekhar}.
As one enters the near field, the changes in the spectrum become
dramatic because surface resonances may appear in the range of
thermal frequencies. Polar materials provide a generic example
where optical phonons hybridize with the electromagnetic field
into surface phonon polaritons. These resonances enhance the
electromagnetic energy density and they also polarize the radiation
field. An example is shown in Fig.\ref{fig:SiC-glass} for a planar
body: at the wavelength $11.36\,\mu{\rm m}$, silicon carbide
has surface resonances (because the permittivity is negative)
which show up as a prominent `shoulder' at distances $d \approx
\lambda / 10$. At even shorter distances, the energy density
follows a law $1/d^2$ which can be explained by the
electrostatic fields of fluctuating dipole moments. The surface
resonance appears as an enhanced amplitude of this power tail.

The polarization of the near field has been quantified in 
the plot by the following degree of polarization%
\footnote{Eq.(\ref{eq:d-of-p})
arises from Refs.\cite{Friberg02,Ellis05} which coincide in
the present situation. They do not so in the general case,
because of different ways the three field components have
been taken into account to define degrees of polarization.
For a comparison, see Ref.\cite{Greffet07b}.
}
\begin{equation}
P = \frac{ |\varepsilon_{\Vert} - \varepsilon_{zz}| }{
\varepsilon_{zz} + 2 \varepsilon_{\Vert} }
\,,
\qquad
0 \le P \le 1
\label{eq:d-of-p}
\end{equation}
where $\varepsilon_{zz}$ ($\varepsilon_{\Vert}$)
is proportional to the spectrum of 
the electric field perpendicular (parallel) to the surface
and by symmetry, 
$\langle E_x^2( {\bf r} ) \rangle_\omega 
= \langle E_y^2( {\bf r} ) \rangle_\omega$. Note the strong
polarization when the surface resonance is excited. In the limits
of short and large distances, a partial polarization $P \approx 1/4$
arises, but for different reasons: at short distances, it follows
from electrostatics that the normal field ($E_z$) has a spectrum
twice as large as the parallel field. At large distances, the
polarization is due to reflection and emission at glancing angles 
from the surface, similar to the Brewster effect. This partial
polarization is actually
an artefact of assuming a planar source of infinite extent. For
a discussion of the thermal emission of a spherical source,
see for example Refs.\cite{Agarwal04b, ZuritaSanchez16}.

Let us mention that to measure the energy density, a local
measurement with a small pick-up antenna is needed. This can
be achieved with sharp tips that scatter the near 
field~\cite{Babuty13},
see the contribution of De Wilde in this issue. The
scattered signal is dominated by the immediate environment
of the local probe: a lateral resolution comparable
to the distance $d$ would be typical. In this way, the thermal
radiation may provide a scanning image of local electromagnetic
properties and resolve metallic nano-objects on a dielectric
substrate, for example.

\subsection*{Energy balance and radiative heat transfer}

The radiative heat flux between two objects is described
in electromagnetism by the Poynting vector.
% ${\bf S} = {\bf E}
%\times {\bf H}$. 
In our context, it has a natural spectral representation and
involves the Planck spectrum as an essential factor, due
to the assumption of local thermal equilibrium of the radiation
sources. The symmetry of photons propagating to the left and
the right is broken by the difference in temperature. 
As the bodies are approached to sub-wavelength distances,
the heat flux is enhanced due to near-field coupling. 

Let us start with the energy conservation law for the 
Maxwell-Langevin equations. In a stationary situation (the 
energy density is constant), one gets 
\begin{equation}
\nabla \cdot 
\mathop{\rm Re} \langle {\bf E} \times {\bf H} \rangle_\omega
+
\omega \mathop{\rm Im}\varepsilon( \omega )
\langle {\bf E}^2\rangle_\omega
=
-
\mathop{\rm Re} \langle {\bf j} \cdot {\bf E} \rangle_\omega 
\label{eq:power-conservation}
\end{equation}
for each frequency $\omega$. The
first term on the lhs involves the radiative emission spectrum
(average Poynting vector $\langle {\bf S} \rangle_\omega$), 
the second term the absorption in the medium. 
(With our sign convention, passive media have
$\omega \mathop{\rm Im}\varepsilon( \omega ) \ge 0$.)
On the
rhs, we find the power transferred to the fields by the 
mechanical motion of the sources~\cite{Jackson}. 
Eq.(\ref{eq:power-conservation}) has been commented upon
in a discussion of energy conservation in a dissipative 
system~\cite{Ford93a}: the electromagnetic energy lost by
absorption in the medium is `replenished' by the sources
in the same medium, as long as local equilibrium holds. This
observation illustrates the general idea of treating dissipative
quantum systems with Langevin dynamics: the fluctuating forces
maintain the system energy at its equilibrium value.
By the same token, they prevent the commutators of the system
observables from decaying.

\begin{figure}[htb]
\centerline{%
\includegraphics*[width=90mm]{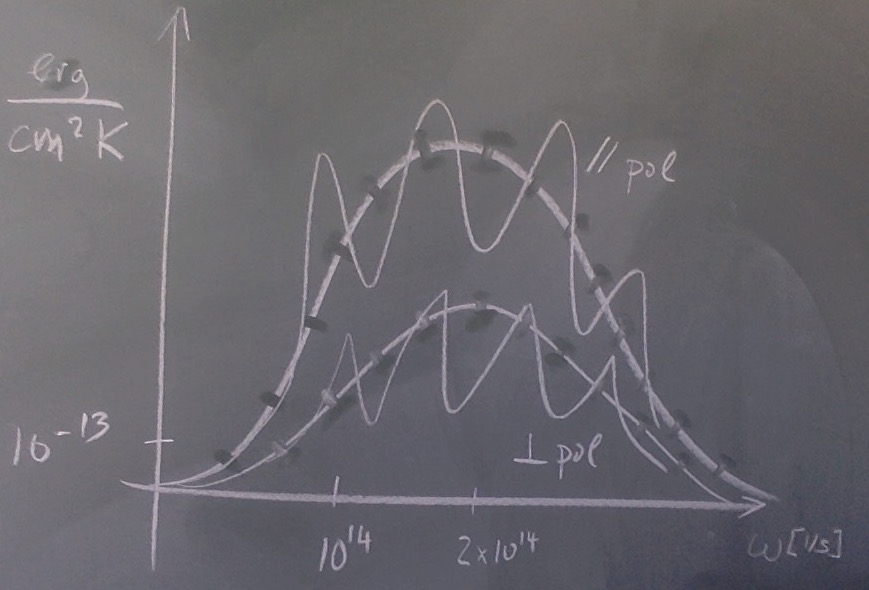}}
\caption[]{Spectrum of heat flux between two metallic slabs,
contribution of propagating waves ($0 \le k \le \omega/c$
in Eq.(\ref{eq:Polder71-flux})). Upper pair of curves: 
electric field in the plane of incidence (p- or TM-polarization),
lower pair: electric field perpendicular to plane of incidence
(s or TE). Solid lines: distance $d = 10\,\mu{\rm m}$, dashed 
lines: $d \to \infty$. The peaks arise when $d$ is an integer
multiple of $\lambda / 2$ (standing waves).
The bodies are at temperatures
$T_h = 315\,{\rm K} = T_c + 1\,{\rm K}$, and have identical
conductivities described by a Drude model with 
$\sigma_{\rm DC} = 3.9 \,{\rm MS/m}$
and damping time $\tau = 7.1\,{\rm fs}$.
\\
Sketch after Fig.~4 of Ref.\cite{Polder71}:
D. Polder and M. van Hove,
``Theory of radiative heat transfer between closely spaced bodies'',
\emph{Phys. Rev. B} {\bf 4} (1971) 3303.
}
\label{fig:Polder71fig4}
\end{figure}

In a typical setting of heat transfer between two bodies,
we may evaluate Eq.(\ref{eq:power-conservation}) in a vacuum gap
between them and get 
$\nabla \cdot \langle {\bf S} \rangle_\omega = 0$. In a planar
geometry, this means that the normal flux 
is constant. We quote
here the formula that results from evaluating the mixed electric 
and magnetic correlation function between two planar bodies,
`hot' and `cold', at distance $d$~\cite{Polder71,Loomis94}
\begin{eqnarray}
\langle {S}_z \rangle_\omega &=& 
\left[ 
\Phi( \omega, T_h)
-
\Phi( \omega, T_c)
\right]
\sum_{p}
\int\!\frac{ k \,{\rm d}k }{ (\omega/c)^2 }
A_{hp} A_{cp}
\left|
\frac{ {\rm e}^{ 2 {\rm i} k_z d } }{
1 - r_{hp} r_{cp}\,{\rm e}^{ 2 {\rm i} k_z d } }
\right|^2
\label{eq:Polder71-flux}
\\
\Phi( \omega, T) &=&
\frac{ \hbar \omega^3 }{ 4\pi^3 c^2 } 
\frac{ 1 }{ {\rm e}^{ \hbar \omega / k_B T } - 1 }
\end{eqnarray}
where $\Phi( \omega, T)$ is the flux of a black body.
Two polarizations $p$ are summed over, and 
the quantity
\begin{equation}
0 \le 
A_{bp} = \left\{
\begin{array}{ll}
1 - |r_{bp}|^2 & \mbox{if $k \le \omega/c$}
\\
2 \mathop{\rm Im} r_{bp} & \mbox{if $k > \omega/c$}
\end{array}
\right.
\label{eq:}
\end{equation}
is proportional to the absorption of a wave incident on
body $b = h, c$ where it is reflected with amplitude 
$r_{bp} = r_{bp}( k, \omega )$. Finally, 
$k_z = \sqrt{(\omega/c)^2 - k^2}$ is the normal
component of the wavevector in vacuum. For $k > \omega/c$,
it is purely imaginary (evanescent wave, $\mathop{\rm Im}
k_z \ge 0$), 
the heat then transports by photon tunnelling
from the bulk of one body to the other.
The denominator $1 - r_{hp} r_{cp}\,{\rm e}^{ 2 {\rm i} k_z d }$
in Eq.(\ref{eq:Polder71-flux})
describes the multiple reflection of waves between the
two interfaces; its zeros correspond to `cavity 
modes' and surface resonances. In practice, due to 
imperfect reflections, the modes are broadened but are
still visible in the spectrum. 
See Fig.\ref{fig:Polder71fig4} for an example, taken
from the seminal paper by Polder and Van Hove\cite{Polder71}.

The enhancement of the heat flux in the near field is mainly
due to the contribution of evanescent waves (photon tunnelling).
Taking the limit of purely non-absorbing bodies, one finds
that the $k$-integral in Eq.(\ref{eq:Polder71-flux}) is limited
to $k \le n \omega /c$ where $n$ is the refractive index: in
this case, the Planck spectrum (Stefan-Boltzmann law) still 
holds with a modified prefactor (angle-averaged emissivity). 
Fig.~\ref{fig:Polder71fig4} 
illustrates that the propagating modes that dominate in this 
range mainly re-distribute the emission spectrum by forming
cavity resonances.
The qualitatively new power laws
seen in Fig.\ref{fig:SiC-glass}
emerge from deeply evanescent waves, $k \gg \omega / c$. 
These correspond, on one hand, to surface polaritons 
with typically $k \sim 1 / d$ 
where $d$ is the distance,
and on the other, to diffusion `modes', 
$k \sim (\mu_0 \sigma_{\rm DC} \omega)^{1/2}$,
where $(\mu_0 \sigma_{\rm DC})^{-1}$ 
is the magnetic diffusion constant
in a conductor. From an analysis of the two polarizations,
these contributions have been pointed out already
by Polder and Van Hove~\cite{Polder71}.

\section*{Challenges}

For recent applications of fluctuation electrodynamics, 
see other papers in this issue. We conclude this discussion
with two challenges that may lead to a refined theory. By
analogy to the unexplained perihelion shift of Mercury
about hundred years ago, we suggest to name the following
observations `anomalies'.

\subsection*{Casimir force and conductivity}

The first anomaly has been dubbed the `plasma vs.\ Drude
controversy' in the community working on dispersion forces
and the Casimir effect~\cite{DalvitBook,SimpsonLeonhardt15}.
We focus here on an experiment, performed by the
Mohideen group~\cite{Chen07b}, that does not seem to have
received much attention in this discussion. 
In the setup, an atomic force
microscope is measuring the
force between a sphere (diameter $\sim 200\,\mu{\rm m}$) 
and a planar body. 
The body is a semiconductor membrane (silicon) 
and gets irradiated with a train of laser pulses
that increases the carrier density. 
The density is practically stationary over the $\sim 5\,{\rm ms}$
duration of the pulses, and the force is measured with a
lock-in amplifier at the repetition rate.
From the viewpoint of macroscopic electrodynamics, 
the irradiation changes the conductivity 
by a few orders of magnitude
from quite small 
(intrinsic, related to defects, $\sim 10\,{\rm S/m}$) 
to high ($\sim 1\,{\rm MS/m}$), 
while the temperature increase is negligible. 

The calculated change in the electromagnetic force agrees 
with the experimental data within $10\%$, 
but only for the laser-doped material. 
A statistically significant difference to the theory
(about $50\%$ at distance $d \approx 100\,{\rm nm}$) 
is observed for the intrinsic Si membrane. 
The deviation can be pinpointed to the behaviour of the material
conductivity across the thermal spectrum
(see Refs.\cite{Chen07a,Svetovoy08b} for a discussion).
Technically speaking, frequency integrals are performed
as summations over the imaginary axis (Matsubara sum), 
using the analytical continuation of response functions. 
The anomaly arises from the zero-frequency term in the sum. 
The current discussion about this thermal Casimir anomaly 
has led to experiments with magnetic materials (nickel) 
where the deviation to theory amounts to 
a few orders of magnitude~\cite{Bimonte16}.
An explanation for this anomaly would turn precision
measurements of forces in the sub-micrometer range 
into tighter constraints on gravity on small scales
(including extensions of the Standard Model 
like the `fifth force' and additional
dimensions)~\cite{Decca07a,Decca07b,Chen16b}.

\subsection*{Heat transfer below $10\,{\rm nm}$}

%Achim Kittel experiments

Two recent experiments on heat transfer between sharp,
metallized tips and a substrate, held at different temperatures,
suggest a similar anomaly when compared to fluctuation 
electrodynamics~\cite{Kim15a,Kloppstech15}. The setups differ
mainly in the size of the tips (radius of curvature). The
group of Reddy and Meyhofer found good agreement with 
theoretical calculations done by the group of
Garc\'{\i}a-Vidal and Cuevas \cite{Kim15a}, within uncertainties
related to surface roughness. Distances down to a few
nanometers were reached (tip radius $\approx 450\,{\rm nm}$). 
The group of Kittel observed a heat flux 
much larger than theory (computed by Biehs and Rodriguez)~\cite{Kloppstech15}, 
using a sharper tip (radius $\approx 30\,{\rm nm}$) 
in a similar range of distances. The data show an onset of
the `giant heat flux' at $d \approx 5\ldots6\,{\rm nm}$, with
a roughly linear increase as $d$ is reduced.
According to calculations, a significant fraction of the heat
flux should originate from the `shaft' of the tip (conical
shape with $\sim 300\,{\rm nm}$ height). It has been checked
that the difference between experiment and theory is not reduced
when a nonlocal conductivity (spatial dispersion) is 
applied~\cite{Kloppstech15}. The solution to this
anomaly in short-range heat transfer is currently under
investigation~\cite{Wang16a,Wang16b}. 
Molecular dynamics simulations like those 
performed by the Volz~\cite{Volz05} 
and Chen groups~\cite{Chiloyan15} 
are likely to
provide a versatile tool to `bridge the gap' between the
mesoscopic and microscopic scales and include, for example,
heat transport from phonons~\cite{Mingo09}.

\section*{Conclusion}

We hope that the present introduction to fluctuation electrodynamics
provides a convenient `entry point' into this powerful method.
As it happens with any other physical theory, it must be
used having its limitations in mind. Some of these are quite
natural and related to convenient approximations like a local
response. The limits set by experiments
that give different results, are pointing towards new challenges 
and may open up directions for further development. 

\subsubsection*{Acknowledgments}

I thank the \emph{Deutsche Forschungsgemeinschaft} for support
through the DIP program (grant number FO-703/2-1).

%====================================
%           Bibiliography           %
%====================================
%
% \newpage

\bigskip

\providecommand{\newblock}{}

\end{document}